\font\eightrm=cmr8 at 8pt
\documentclass{oxarticle}
\usepackage{amsmath, amssymb}
\usepackage{graphics, setspace}
\usepackage{epstopdf}
\usepackage{bm}
\usepackage{tikz}
\usepackage{xcolor}

\newcommand{\XM}{{Exotic Model}}

\newcommand{\ME}{Master Equation}

\newcommand{\cM}{{\cal M}}

\newcommand{\CDSS}{{Chiral Dotted Spinor Superfield}}

\newcommand{\cA}{{\cal A}}

\newcommand{\Exists}{\bm{\exists}\kern-0.6em\bm{\exists}}

\newcommand{\EI}{Exotic Invariant}
\newcommand{\ei}{exotic invariant}

\newcommand{\SSM}{Supersymmetric Standard Model}

\newcommand{\sg}{$\rm SU(3) \times SU(2) \times U(1)$}

\newcommand{\bsg}{$\rm SU(3)  \times U(1)$}

\newcommand{\bt}{\begin{tabular}{c}}
\newcommand{\et}{\end{tabular}}

\newcommand{\eb}{\ee\be } 
\newcommand{\ebp}{\rt.\ee\be\lt.} 
\newcommand{\ebpp}{\rt.\rt.\ee\be\lt.\lt.} 
 
\newcommand{\bmat}{\lt ( \begin{array} }
\newcommand{\emat}{  \end{array} \rt )}

\newcommand{\oP}{{\ov P}}

\newcommand{\oE}{{\ov E}}

\newcommand{\oY}{{\ov Y}}

\newcommand{\oy}{{\ov \y}}

\newcommand{\oC}{{\ov C}}

\newcommand{\oF}{{\ov F}}
\newcommand{\A}{{\ov A}}

\renewcommand{\a}{\alpha}	
\renewcommand{\b}{\beta}
\newcommand{\g}{\gamma}
\renewcommand{\d}{\delta}

\newcommand{\z}{\zeta}
\newcommand{\h}{\eta}

\newcommand{\lam}{\lambda}
\newcommand{\m}{\mu}
\newcommand{\n}{\nu}	
\newcommand{\x}{\xi}

\newcommand{\s}{\sigma}

\newcommand{\f}{\phi}
\renewcommand{\c}{\chi}
\newcommand{\y}{\psi}
\newcommand{\w}{\omega}
\newcommand{\G}{\Gamma}
\newcommand{\D}{\Delta}
	
\newcommand{\Lam}{\Lambda}

\renewcommand{\S}{\Sigma}

\newcommand{\la}{\label}
\newcommand{\ci}{\cite}

\newcommand{\ds}{\documentstyle}	
\newcommand{\fr}{\frac}

\newcommand{\pa}{\partial}
\newcommand{\ov}{\overline}
\newcommand{\br}{\begin{rant}}
\newcommand{\er}{\end{rant}}
\newcommand{\beC}{\begin{Conjecture}}
\newcommand{\eeC}{\end{Conjecture}}
\newcommand{\be}{\begin{equation}}
\newcommand{\ee}{\end{equation}}
\newcommand{\ba}{\begin{array}} 
\newcommand{\ea}{\end{array}}
\newcommand{\bea}{\begin{eqnarray}}
\newcommand{\eea}{\end{eqnarray}}
\newcommand{\ra}{\rightarrow}

\newcommand{\Lra}{\Leftrightarrow}
\newcommand{\lt}{\left}
\newcommand{\rt}{\right}

\newcommand{\ben}{\begin{enumerate}}
\newcommand{\een}{\end{enumerate}}
\newcommand{\bitem}{\begin{itemize}}
\newcommand{\eitem}{\end{itemize}}

\newcounter{orange} 
\newcounter{apple} 
\addtocounter{apple}{1}
\newcounter{grape} 
\addtocounter{grape}{1}
\newcommand{\numberhere}{Feb5\;}
\newcommand{\articlenumber}{E5-FINAL}

\proofmodefalse
\usepackage{color}

\newcommand{\mathsym}[1]{{}}
\newcommand{\unicode}[1]{{}}

\usepackage{tikz}
  
\begin{document}

 \begin{center}
{\huge The EP Model with U(1) (E5)}
\\
[1cm]

\renewcommand{\thefootnote}{\fnsymbol{footnote}}

{ John A. Dixon\footnote{jadixg@gmail.com,
john.dixon@ucalgary.ca}\\Physics Dept\\University of Calgary
\\[1cm]}  
 \end{center}

\Large
 
 \begin{center}Abstract
 \end{center}
 
Here we add a U(1) gauge theory to the simple EP exotic
invariant model in the paper E4. This paper E5 is the fifth in a
series of papers En.

 \section{Introduction} 
\Large

This paper is a preparation for the paper E6, which introduces
the Exotic Model, which is a combination of the \SSM\ with the
Exotic Invariant that couples to it. This paper describe the
same simple model as E4, but we add a U(1) gauge theory here,
and note the necessary changes.

\subsection{Field parts of the action}
\la{firstpartofaction}

This section follows the same pattern as the similar sections in
\ci{E3} and \ci{E4}.
Here we will assemble the action for the matter and the gauge
theory in this model.
The similarity to the terms in \ci{E3} and \ci{E4} is very
evident:

Here is the total action:
\be
\cA= \cA_{\rm Fields}+\cA_{\rm PseudoFields}
\la{sumfofieldandpseudo}
\ee

\subsection{The Action $\cA_{\rm Fields}$}
As in \ci{E4}, we have the following
\be
\cA_{\rm Fields\;U(1)}= {\cal A}_{\rm E\;U(1)} + {\cal A}_{\rm
P\;U(1)} +\cA_{\rm CDSS,\;Kinetic} \eb
+{\cal A}_{\rm Mass } +{\cal \A}_{\rm Mass }  
+\cA_{\rm CDSS,\;Chiral} +{\cal \A}_{\rm CDSS,\;Chiral} +
{\cal A}_{\rm  SUSY\;Gauge\;  U(1) }
 \ee
But now there are extra terms in the following parts, because we
include covariant derivatives for the fields here. The covariant
derivatives can be found in the Table (\ref{brstransE}) below:
\be {\cal A}_{\rm E\;U(1)} = 
\int d^{4}x \left \{
  F_E    {\ov F}_{E}
  -
\y_E^{ \a  }  D^{\a \dot \b}{\ov \y}_{ E}^{  \dot \b}  
-   D^{\a \dot \b}
  E  \ov D_{\a \dot \b}  \oE  
\ebp
  + g_1 D' E \ov E
  +i  g_1 \y_E^{ \a  } \lam_{ \a  } \oE
-i  g_1 {\ov \y}_{ E}^{  \dot \b}  {\ov \lam}_{  \dot \b}    E
\rt\}
\la{Ekineticaction2}\ee

\be {\cal A}_{\rm P\;U(1)} = 
\int d^{4}x \left \{
  F_P    {\ov F}_{P}
  -
\y_P^{ \a  }  D^{\a \dot \b}{\ov \y}_{ P}^{  \dot \b}  
-   D^{\a \dot \b}
  P  \ov D_{\a \dot \b}  \oP  
\ebp
 -  g_1 D' P \ov P
  -i  g_1 \y_P^{ \a  } \lam_{ \a  } \oP
+i  g_1 {\ov \y}_{ P }^{ \dot \b}  {\ov \lam}_{  \dot \b}   P
\rt\}
\la{Pkineticaction2}\ee
The terms $
{\cal A}_{\rm Mass } +{\cal \A}_{\rm Mass }  
+\cA_{\rm CDSS,\;Chiral} +{\cal \A}_{\rm CDSS,\;Chiral} $ are
exactly the same here as they were in E4 (\ci{E4}) and we will
not repeat them here, since it is essential to read E4 before
reading this paper. But we also have the following gauge action
which is new to this paper E5:

 \be 
{\cal A}_{\rm  SUSY\;Gauge\;  U(1) } = 
\eb
\int d^{4}x \left \{ -\frac{1}{4} 
\lt (\pa_{\m} V_{\n} 
-\pa_{\n} V_{\m} \rt ) \lt ( \pa^{\m} V^{ \n} 
-\pa^{\n} V^{ \m} \rt )
 - \frac{1}{2}  \lambda^{ \a}
 \pa_{\a \dot \b}  {\ov  \lambda}^{ \dot \b} 
 + \frac{1}{2}
D D  \rt \}
\ee

In the above $ V^{ \n}$ is a vector field. We will use the
equivalent form
\be
V_{\a \dot \b} =c \s^{\n}_{\a \dot \b} V_{\n}
\ee  
where the well known Pauli matrices $\s^{\n}_{\a \dot \b}$ will
be discussed again in \ci{E10} and c is a constant that we will
discuss later.
We also have a new auxiliary scalar field D, and a new spinor
field
 $\lambda^{ \a}$, along with its complex conjugate 
$ {\ov \lambda}^{ \dot \a} $. There are also three other new
fields that are in this multiplet, though they are not in this
action ${\cal A}_{\rm SUSY\;Gauge\; U(1) }$. We will see them
below in (\ref{varofu1}). These are the ghost $\w$ and antighost
$\h$ which are Grassmann odd, and the auxiliary Z which is
Grassmann even.

\subsection{PseudoField parts of the action}

Now we assemble the pseudofield action for the matter in this
model:

\be
\cA_{\rm PseudoFields}= {\cal A}_{\rm PseudoFields,E} + {\cal
A}_{\rm PseudoFields,P }
+\cA_{\rm PseudoFields,CDSS}+ * \eb +{\cal A}_{\rm
PseudoFields\; U(1)}+ {\cal A}_{\rm Structure} \ee
where the following have some modifications from the presence of
the U(1) coupling. The ghost $\w$ appears in these terms with a
coupling $g_1$, and there are also terms with $ g_1 {\ov
\lambda}^{ \dot \b} $, which are necessary here to provide
invariance of the action (\ref{sumfofieldandpseudo}), and
nilpotence for the BRS transformations:
\be
{\cal A}_{\rm PseudoFields, E}  = \int d^4 x \;
\lt\{ \G_{E} \lt ( \y_{ E \b} { C}^{ \b} + i g_1 \w E+ \x^{\nu}
\partial_{\nu} E \rt )
\la{e1}
\ebp +
 Y_{E}^{ \a}
\lt ( \pa_{ \a \dot \a } E \oC^{\dot\a} + i g_1 \w \y_{ E \a}
 + F_{E} {C}_{ \a}
+ \x^{\nu} \partial_{\nu}  \y_{ E \a}
\rt )
\ebp
+
\Lam_{E} 
 \lt (
D_{\a \dot \b} \y_E^{ \a} {\ov C}^{\dot \b} + i g_1 \w F_E+ i g_1 {\ov
\lambda}^{ \dot \b} {\ov C}_{ \dot \b} E
+ \x^{\g \dot \d} \partial_{\g \dot \d}  F_E 
\rt )
\rt \}
\la{e3}
\ee
\be
{\cal A}_{\rm PseudoFields, P} = \int d^4 x \;
\lt \{
 \G_{P}    \lt ( \y_{ P \b} {  C}^{  \b} - i  g_1  \w    P 
+ \x^{\nu} \partial_{\nu}  P\rt )
\la{p1}\ebp
+
 Y_{P}^{ \a}
 \lt ( \pa_{ \a \dot \a  }  P-  i  g_1  \w    \y_{ P \a} 
 + F_{P} {C}_{ \a}
+ \x^{\nu} \partial_{\nu}  \y_{ P \a}
\rt )
\la{p2}\ebp
+
\Lam_{p}  \lt ( {\ov C}^{\dot \a}
 \pa_{ \a \dot \a  }  {  \y}_{P}^{  \a} -  i g_1  \w  F_P
- i  g_1 {\ov \lambda}^{ \dot \b} {\ov C}_{ \dot \b}   P
 + \x^{\nu} \partial_{\nu}   F_P   
\rt )
\rt \}
\la{p3}\ee
As in \ci{E4} we still have the unmodified form:
\be
{\cal A}_{\rm PseudoFields, CDSS} = \int d^4 x \;
\lt \{ G^{\dot\a}\lt (  C^{\a} W_{\dot \a \a}+
 \x^{\g \dot \d} \partial_{\g \dot \d}\f_{\dot \a}\rt )
 \ebp
+\S^{ \dot \b \a}\lt (
\pa_{ \a \dot \g }  \f_{\dot\b} {\ov C}^{\dot \g}  
+ 
C_{\a}   
\c_{\dot\b}
+ \x^{\g \dot \d} \partial_{\g \dot \d}  W_{\dot \b \a}
\rt )
+L^{\dot \a}\lt (
  \pa^{\b \dot \g}  W_{ \dot \a \b}   {\ov C}_{\dot \g} 
+ \x^{\g \dot \d} \partial_{\g \dot \d}  \c_{\dot\a}
\rt ) \rt \}
 \ee
\be
 {\cal A}_{\rm Structure} =
 -X_{\a \dot \b} C^{\a} \oC^{\dot \b}
\ee
Here we need the following new ${\cal A}_{\rm PseudoFields\;
U(1)}$:

\be
{\cal A}_{\rm PseudoFields\; U(1)} = \int d^4 x \;
\left \{ 
\S^{ \a \dot \b } 
\lt ( 
  \pa_{\a \dot \b} \w^{}
  + 
\fr{1}{2} C_{\a} 
\ov{\lambda}^{}_{ \dot{\b} }   +
\fr{1}{2} \lambda^{ }_{\a } 
{\ov C}_{\dot{\b} }   + \x^{\nu}
\partial_{\nu} V_{\a \dot \b}^{} \rt )
\ebp
+
 L^{ \a} 
  \lt (
   \frac{1}{2}
 \lt ( \pa_{\m} V_{\n} 
-\pa_{\n} V_{\m} 
 \rt )\s^{\mu \nu}_{\a \b} C^{\b}     + i D^{}
C_{\a}  + \x^{\nu} \partial_{\nu}
\lambda^{}_{\a} 
 \rt )
	\la{variationoflambda}
		\ebp 
+ \ov{L}^{ \dot \a}  
\lt (  \frac{1}{2}
 \lt ( \pa_{\m} V_{\n} 
-\pa_{\n} V_{\m} 
 \rt ) \ov{\s}^{\mu \nu}_{\dot{\a} 
\dot{\b} } 
\ov{C}^{\dot \b}      -  i D^{}  {\ov C}_{ \dot \a}  +
\x^{\nu} \partial_{\nu}\ov{
\lambda}^{}_{\dot \a}  
\rt )
\ebp
+
W^{}
\lt (
  C^{\a} \ov{C}^{\dot \b}  V_{\a
\dot \b} +
\x^{\nu} \partial_{\nu} \omega^{}\rt )
+
H^{} 
\lt ( Z + \x^{\nu} \partial_{\nu}  \h 
\rt )
\ebp+
\z^{}
\lt ( C^{\a} \s^{\mu}_{\a \dot{\b} }
\ov{C}^{\dot \b}
\pa_{\m} \h + \x^{\m} \pa_{\m} Z 
\rt )
\ebp
+
\D^{}
\lt (\fr{-i}{2}  C^{\a} 
 \pa_{\a \dot \b}  {\ov  \lambda}^{ \dot \b} 
 +\fr{i}{2}\ov{C}^{\dot \b}
  \pa_{\a \dot \b}  { \lambda}^{ \a} 
   +
\x^{\nu} \partial_{\nu} D 
\rt )
\right \}
\la{varofu1}
\ee
In ${\cal A}_{\rm PseudoFields\; U(1)}$ there are a number of
new fields. To start with we have the various pseudofields shown
above. This can be written in the form
\[
{\cal A}_{\rm PseudoFields\; U(1)} = \int d^4 x \;
\left \{ 
\S^{ \a \dot \b } 
\lt ( \d V_{\a \dot \b}^{} \rt )
 +
 L^{ \a} 
  \lt (
\d
\lambda^{}_{\a} 
 \rt )
+ \ov{L}^{ \dot \a}  
\lt ( \d \ov{
\lambda}^{}_{\dot \a}  
\rt )
\rt. \]
\be
\lt.
+
W^{}
\lt (
 \d \omega^{}\rt )
+
H^{} 
\lt (\d  \h 
\rt )
+
\z^{}
\lt ( \d Z 
\rt )
+
\D^{}
\lt (\d D 
\rt )
\right \}
\la{varofu2}
\ee
Note in the above expressions, we also have the new Faddeev
Popov Abelian ghost field $\w$ and the gauge auxiliary field Z,
and the Faddeev Popov Abelian antighost field $\h$. The ghost
$\w$ and antighost $\h$ are Grassmann odd, and the auxiliary Z
is Grassmann even. As usual the pseudofields have the opposite
Grassmann parity to the fields that are dual to them. The ghost,
antighost and auxiliary are necessary to form the Gauge Fixing
and Ghost action, but in this paper we do not need to discuss
those.

\section{The Exotic Invariant for the U(1) Case}

Now we are ready to add the exotic invariant here. It is
essentially the same as we found in E4, except that the terms
have new pieces in them:
\be
\cA_{\rm X, U(1) } = \cA_{\rm X, E, U(1) } -\cA_{\rm X, P, U(1)
}
\la{exoinvhere}
\ee
The mechanism that solves the constraint equation here is the
same as it was in E4.

The first term in (\ref{exoinvhere}) 
is given by
\be
\cA_{\rm X, E, U(1) } = \int d^4 x \lt \{ \f^{\dot \a}\lt [ \lt
(
b_{1}E \G_{E} 
+ b_{2}\y_{E}^{\b} Y_{E \b}
+b_{3}\Lam_{E}  F_{E}
\rt ) \oC_{ \dot \a}
\la{eiforE1}
\ebpp + 
b_{4}  \y_{E}^{\a} \ov D_{\a \dot \a} \oE + 
b_{4} g_1 \ov \lam_{\dot \a} E \oE 
 + 
b_{5} F_{E}\oy_{E  \dot \a}\rt ]
\ebp
+ W^{\dot \a \b} \lt [ \lt (b_{6} E Y_{E \b} + b_{7} \y_{E \b}
\Lam_{E}\rt )
\oC_{ \dot \a}+ 
b_{8} \y_{E \b}\oy_{E \dot \a} + 
\ebpp b_{9}{E}\ov D_{\b \dot \a} \oE   \rt ]
  +  \c^{\dot \a} \lt [b_{10}  E \Lam_{E } \oC_{ \dot \a} +
b_{11}E\oy_{E\dot \a}
 \rt ] \rt \}  
\la{eiforE4}  
\ee
and
\be
b_{1}
=
b_{2}=
b_{3}=
b_{4}=b_{7}=b_{10}=1; 
b_{5}=
b_{6}=
b_{8}=
b_{9}=
b_{11}= -1
\la{valsofb}
\ee

The coefficients $b_{i}$ are determined by the requirement that\be
\d ( \cA_{\rm X, E, U(1) } -\cA_{\rm X, P, U(1) })=0
\ee
where the nilpotent antiderivations $\d$ can be found in section
\ref{transformsec}. These are the same as they were in E4, but
there are some new terms in (\ref{eiforE1})--(\ref{eiforE4})
that arise because of the gauge interactions. We need to use the
covariant derivative $\ov D_{\a \dot \a} \oE$, which can be
found below in Table (\ref{brstransE}), and the new term $b_{4}
g_1 \ov \lam_{\dot \a} E \oE $ also plays an important role.

The variation here is the same as it was in E4, namely: \be
\d  \cA_{\rm X, E, U(1) }=  
\d \cA_{\rm X, P, U(1) }
\la{variationsofXEandXP1}
\eb
-\int d^4 x \lt \{
   -  \f^{\dot \a}   E   m  F_P    \oC_{ \dot \a} 
 - \f^{\dot \a}  F_{E}  \oC_{ \dot \a} m P  
-X^{ \a\dot \a}    E \oC_{ \dot \a}m
\y_{P \a} 
\ebp
 - X^{ \b\dot \a}   \y_{E \b} \oC_{ \dot \a} m P  
+   \f^{\dot \a} \y_{E}^{\a} \oC_{ \dot \a}m
\y_{P \a} 
-  \c^{\dot \a}    E \oC_{ \dot \a} m P  
\rt \}
\la{variationsofXEandXP2}\ee
This expression is symmetric when $E\Lra P$ and so, for the
difference, we get the same equation that we had in the paper
E4:
\be
\d \cA_{\rm X,U(1)} =\d \lt \{\cA_{\rm X, E, U(1)} - \cA_{\rm X,
P, U(1) } \rt \}=0
\ee 
This does not change from E4 because it does not contain any
derivatives, and so the gauge theory has very little effect on
the SUSY cohomology here, as will be shown more formally in
\ci{E??}.
The same situation obtains when we add other gauge couplings
such as couplings to non-Abelian gauge theories, as we need to
do for the XM in E6. So we note that the situation does not
change much when we have gauge invariance, except that there are
important new terms in the exotic invariants.

\section{The Completion Terms for the Special Exotic Invariant
with U(1), generalized from the EP Case}
\la{compterms}
In E4 we showed that it was necessary to complete the action
with completion terms. These are the same here as they were
there, and so we will not repeat them here. The basic reason
that they do not change is that there are no derivatives in the
Completion Terms.

\section{The \ME}
\la{mastereqsection}

Now the  \ME\ has the following form: 
\be
\cM_{\rm }=\cM_{\rm E}+\cM_{\rm P} 
+\cM_{\rm X}  +\cM_{\rm U(1)} +\cM_{\rm Structure}=0
\ee
Here are the detailed forms:
\be
\cM_{\rm E}=
\int d^4 x \lt \{
\fr{\d \cA}{\d E} \fr{\d \cA}{\d \G_{E}}
+
\fr{\d \cA}{\d \y_{E \a}  } \fr{\d \cA}{\d Y_{E}^{\a}}
+
\fr{\d \cA}{\d F_E} \fr{\d \cA}{\d \Lam_{E}}
+
\fr{\d \cA}{\d \oE} \fr{\d \cA}{\d \ov\G_E}
\ebp+
\fr{\d \cA}{\d \oy_{E \dot \a}} \fr{\d \cA}{\d \oY_E^{\dot \a}}
+
\fr{\d \cA}{\d \oF_{E}} \fr{\d \cA}{\d \ov\Lam_E}
\rt \}  
\la{wzh}
\ee
The form $\cM_{\rm P}$ is obtained from $\cM_{\rm E}$ by simply
taking $E\ra P$.
\be
\cM_{\rm X} =\int d^4 x \lt \{
\fr{\d \cA}{\d \f_{\dot\a}} \fr{\d \cA}{\d  G^{\dot\a}}
+
\fr{\d \cA}{\d X_{\a\dot\a}} \fr{\d \cA}{\d \S^{\a\dot\a}}
+
\fr{\d \cA}{\d\c_{\dot\a}} \fr{\d \cA}{\d L^{\dot\a}}
 +\fr{\d \cA}{\d \ov \f_{\a}} \fr{\d \cA}{\d  \ov G^{ \a}}
\ebp+
\fr{\d \cA}{\d \ov X_{\dot\a\a}} \fr{\d \cA}{\d
\ov\S^{\dot\a\a}}
+
\fr{\d \cA}{\d\ov\c_{\a}} \fr{\d \cA}{\d \ov L^{\a}}
\rt \}  
\la{mastercdssdownfield} 
\ee

\be
\cM_{\rm U(1)}= \int d^4 x \; \lt \{
\fr{\d \cA}{\d \S^{\a\dot\b}} \fr{\d \cA}{\d V_{\a \dot \b}} 
+ \fr{\d \cA}{\d  \D } \fr{\d \cA}{\d D} 
+ \fr{\d \cA}{\d  Z } \fr{\d \cA}{\d \z} 
+ \fr{\d \cA}{\d H } \fr{\d \cA}{\d \h}+ \fr{\d \cA}{\d W}
\fr{\d \cA}{\d \w} \rt \}
\la{MU1}\ee

\be
 \cM_{\rm Structure}=
\fr{\pa \cA}{\pa P_{\a \dot \b}} 
\fr{\pa \cA}{\pa \x^{\a \dot \b}}    
\la{masterstructure}
\ee
\section{Tables of the Transformations}
\la{transformsec}

We obtain $\d$ as the square root of the \ME, as usual. 
\be
\la{brstransE} 
\vspace{.1in}
\framebox{{$\begin{array}{lll}  
& &{\rm Nilpotent  \;Transformations\; for\; E\;With \;U(1)}\\
\d E&= & 
\fr{\d {\cal A}}{\d \G_E} 
=  \y_{E  \b} {C}^{  \b} + i g_1  \w E
+ \x^{\g \dot \d} \partial_{\g \dot \d} E
\\
\d {\ov E} &= & 
\fr{\d {\cal A}}{\d {\ov \G}_E} 
=  {\ov \y}_{E  \dot \b} {\ov C}^{ \dot  \b}  - i g_1  \w \oE
+ \x^{\g \dot \d} \partial_{\g \dot \d} {\ov E}
\\

\d \y_{E \a} &  =& \fr{\d {\cal A}}{\d {  Y}_E^{   \a} } = 
D_{ \a \dot \b } E {\ov C}^{\dot \b}  
+  i g_1  \w \y_{E \a}+
C_{\a}   
F_E
+ \x^{\g \dot \d} \partial_{\g \dot \d}  \y_{E\a  }
\\

\d
 {\ov \y}_{E \dot \a} &  =& 
\fr{\d {\cal A}}{\d { {\ov Y}}_E^{i\dot   \a} } = 
\ov D_{ \a \dot \a }  {\ov E}_{} {C}^{\a}  
-  i g_1\w {\ov \y}_{E \dot \a}
+ 
{\ov C}_{\dot \a}   
{\ov F}_{E}
+ \x^{\g \dot \d} \partial_{\g \dot \d} 
 {\ov \y}_{ \dot \a} 
\\
 \d F_E 
&=&\fr{\d {\cal A}}{\d \Lam_E} = 
D_{\a \dot \b} \y_E^{ \a} {\ov C}^{\dot \b} + i g_1 \w F_E+ i g_1 {\ov
\lambda}^{ \dot \b} {\ov C}_{ \dot \b} E
+ \x^{\g \dot \d} \partial_{\g \dot \d}  F_E 
\\
 \d \oF_E 
&=&\fr{\d {\cal A}}{\d \ov\Lam_E} = 
\ov D_{\b \dot \a} \oy_{E}^{\dot \a} { C}^{\b} - i g_1 \w \oF_E
- i g_1 { \lambda}^{ \b} { C}_{ \b} \oE + \x^{\g \dot \d}
\partial_{\g \dot \d} \oF_E
\\
& &{\rm Covariant  \;Derivatives\; With \;U(1)}
\\
D_{\a \dot \b}E & \equiv & \pa_{\a \dot \b} E - i g_1 V_{\a \dot
\b} E;\;
\ov D_{\a \dot \b} \oE \equiv \pa_{\a \dot \b} \oE + i g_1 V_{\a
\dot \b} \oE
\\
D_{\a \dot \b} \y_E^{\a} &\equiv& \pa_{\a \dot \b} \y_E^{\a} - i
g_1 V_{\a \dot \b} \y_E^{\a}
;\;\ov D_{\a \dot \b} \oy_E^{\dot\b} \equiv \pa_{\a \dot
\b}\oy_E^{\dot\b} + i g_1 V_{\a \dot \b}\oy_E^{\dot\b}
\\
\end{array}$}} 
\ee
Here we will not repeat the formulae in E4, and the table of
$\d$ for the U(1) supermultiplet can be read off $\cM_{\rm
U(1)}$ in (\ref{MU1}). But we did write out the new
transformations for the E multiplet in accord with
(\ref{e1})--(\ref{e3}). There is a similar table for $E\ra P$ in
which we must take $g_1 \ra - g_1$, in accord with
(\ref{p1})--(\ref{p3}).

\section{Conclusion}

The three papers E3, E4, E5 briefly summarize the basic facts
about the \EI s, which were introduced informally in E1, and
then deduced using the spectral sequence in E2 . Let us review
the results once more here:
\ben

\item
The full \EI\ in (\ref{exoinvhere}) contains more than one \EI\
type expression, of the form (\ref{eiforE1})--(\ref{eiforE4}),
and they are subtracted, so that their variations cancel each
other as shown in (\ref{variationsofXEandXP1}).
\item
There are a few new terms when we add a gauge interaction with a
gauge theory, but that does not change the basic features of the
\EI\ very much.
\item
It is easy to see that the results here generalize to
non-Abelian gauge groups in the usual way familiar in SUSY
theories.
We need to use covariant derivatives $\ov D_{\a \dot \a} \oE$
and there are new terms like $b_{4} g_1 \ov \lam_{\dot \a} E \oE
$, as we saw in (\ref{eiforE1}). These new terms are very
important, for mass splitting, as we will see in E6, when we
look at the H and K fields and their \EI s.
\item We need to add Completion Terms as in section
(\ref{compterms}) to get the action including the \EI s to
satisfy the original \ME. This is not changed by the presence of
gauge couplings, because these terms have no derivatives.
		\een

These \ei s are now going to be used in E6 to modify the \SSM.
The sum of the special \EI\ and the SSM we will call the XM. It
has a natural gauge symmetry breaking from \sg\ to \bsg. But, as
usual, there is no spontaneous breaking of supersymmetry
included in that gauge symmetry breaking, so the supersymmetry
remains after gauge symmetry breaking. The supermultiplets are
still there.

However the XM contains a number of new and peculiar terms in
the \EI s that contain the H and K fields. In E7 we will deduce
the form of the quadratic ZX action which arises in the XM when
the gauge symmetry breaks from \sg\ to \bsg\ by getting VEVs for
the H and K multiplets. That quadratic ZX action gets mixed with
the Higgs and Photon multiplets by some counterterms that are
needed at one loop as will be shown in E9. Then in E10 we will
show that a computer program confirms the results shown here and
in E6. The remaining unknown, which is very important, and which
has not yet been completed, is to find out whether the tachyons
in these theories can be eliminated by a choice of the
coefficients like the $a_i$ that we used in front of the terms
for the action for the paper E4. This possiblity is suggested by
the results of E7 and E8. That question should also be resolved
in E10.

\begin{center}
 { Acknowledgments}
\end{center}
\vspace{.1cm}

I thank Howard Baer, Friedemann Brandt, Philip Candelas, Mike
Duff, Sergio Ferrara, Dylan Harries, Marc Henneaux, D.R.T.
Jones, Olivier Piguet, Antoine van Proeyen, Peter West and Ed
Witten for stimulating correspondence and conversations. I also
express appreciation for help in the past from William Deans,
Lochlainn O'Raifeartaigh, Graham Ross, Raymond Stora, Steven
Weinberg, Julius Wess and Bruno Zumino. They are not replaceable
and they are missed. I also thank Ben
Allanach, Doug Baxter, Margaret Blair, Murray Campbell, David
Cornwell, Thom Curtright, James Dodd, Richard Golding, Chris T.
Hill, Davide Rovere, Pierre Ramond, Peter Scharbach, Mahdi
Shamsei, Sean Stotyn, Xerxes Tata and J.C. Taylor, for recent,
and helpful, encouragement to carry on with this work. I also
express appreciation to Dylan Harries and to Will, Dave and
Peter Dixon, Vanessa McAdam and Sarunas Verner for encouraging
and teaching me to use coding. I note with sadness the recent
passing of Carlo Becchi and Kelly Stelle, both of whom were
valuable colleagues and good friends.

 \tiny 
\articlenumber\\
\today
\hourandminute


\begin{thebibliography}{99}



\bibitem{bagger} 
J.~Wess and J.~Bagger,
``Supersymmetry and supergravity,''
Princeton University Press, 1992,
ISBN 978-0-691-02530-8

\bibitem{WZ} 
J.~Wess and B.~Zumino,
``Supergauge Transformations in Four dimensions '',
Nucl. Phys. B \textbf{70}, 39-50 (1974)


 

\bibitem{xerxes} 
  H.~Baer and X.~Tata,
``Weak scale supersymmetry: From superfields to scattering
events'',
  Cambridge, UK: Univ. Pr. (2006)  


 
 





 

\bibitem{Weinberg3} Steven Weinberg: ``The Quantum Theory of
fields" Volume 3, Cambridge University Press, ISBN 052155002.
This contains a readable brief history, a summary of the
problems with SUSY, and a useful summary dealing with the
renormalization group in the context of SUSY GUT.





\bibitem{Allanach:2024suz}
B.~Allanach and H.~E.~Haber,
``Supersymmetry, Part I (Theory),'' (Particle Data Group)
[arXiv:2401.03827 [hep-ph]].

\bibitem{D’Onofrio} M. D’Onofrio  and F. Moortgat,  
 ``Supersymmetry, Part II (Experiment)'' (Particle Data Group)
Revised August 2023  

 
\bibitem{haberetal} Dreiner, Herbi K., Howard E. Haber, and
Stephen P. Martin, ``From Spinors to Supersymmetry'' (Cambridge:
Cambridge University Press, 2023)
 

\bibitem{holes} 
J.~A.~Dixon,
``Supersymmetry is full of holes'',
Class. Quant. Grav. \textbf{7}, 1511-1521 (1990)
doi:10.1088/0264-9381/7/8/026



\bibitem{holescommun} 
Ibid. ,
``BRS cohomology of the chiral superfield'',
Commun. Math. Phys. \textbf{140}, 169-201 (1991)


\bibitem{dixminram}  
J.~A.~Dixon, R.~Minasian and J.~Rahmfeld,
"Higher spin BRS cohomology of supersymmetric chiral matter in D
= 4'',
Commun. Math. Phys. \textbf{171}, 459-474 (1995)
[arXiv:hep-th/9308013 [hep-th]].





\bibitem{dixspecseq} 
J.~A.~Dixon,
``Calculation of BRS cohomology with spectral sequences'',
Commun. Math. Phys. \textbf{139}, 495-526 (1991)

\bibitem{dixmin}   
J.~A.~Dixon and R.~Minasian,
``BRS cohomology of the supertranslations in D = 4",
Commun. Math. Phys. \textbf{172}, 1-12 (1995)
[arXiv:hep-th/9304035 [hep-th]]. Some errors of notation have
crept into the appendix to this paper. They should be corrected.
The result is correct, however.




{ The E Series: }The following papers, about exotic invariants,
form a series starting at \ci{E1}, which is E1. They are
labelled (En), where $n=1,2\cdots$.



\bibitem{E1}   
J.~A.~Dixon, `` Supersymmetry anomalies, exotic pairs and the
supersymmetric standard mode (E1)", [arXiv:2407.13673].


 \bibitem{E2}     Ibid. 
``The BRS Cohomology of the Wess Zumino Chiral Scalar
supersymmetric model with exotic pairs and exotic triplets
(E2)'',
[arXiv:2507.14174]



 \bibitem{E3}     Ibid. 
  ``The  simplest Exotic Invariant  (E3)'',
arXiv:2602.01407v1 [hep-ph] 1 Feb 2026

\bibitem{E4}Ibid. 
``The EP Model with Completion Terms (E4)'', arXiv:2602.04501v1
[hep-ph] 4 Feb 2026

  
\bibitem{E5}Ibid. ``The EP Model with U(1) (E5)'', Preprint
issued in February 2026.



\bibitem{E6}Ibid. ``The Supersymmetric Standard Model, combined
with a special \EI, yields a new kind of SUSY mass splitting
(E6)." [arXiv:2507.14381].


  



\bibitem{E7}Ibid. ``The Free Massive Quadratic Action from the
\XM. (E7)'' [arXiv, hep-ph: 2508.06252]




\bibitem{E8} Ibid. ``Removal of the Tachyons from the Fermionic
Sector of the Quadratic ZX Action for the Exotic Model
(E8)"    [arXiv hep-ph: 2512.08561] 


    \bibitem{E9} Ibid. ``More Details for the XM
(E9)"      Preprint to be issued in early 2026.

\bibitem{E10} Ibid. ``Some Mathematica Nfor Exotic Invariant
Models (E10)", Preprint to be issued in early 2026.




\bibitem{E??}Ibid. 
``The BRS cohomology of SUSY Including Gauge Theory and the
\CDSS\ (E??)''. Preprint to be issued, but not soon.



\end{thebibliography}
\end{document}